\documentclass[a4paper,UKenglish,cleveref, autoref, thm-restate]{lipics-v2021}

\usepackage{booktabs}
\usepackage{makecell}
\usepackage{siunitx}
\usepackage[linesnumbered,lined]{algorithm2e}

\usepackage{tikz}
\usetikzlibrary{calc,positioning,decorations.pathreplacing,backgrounds,math,shapes,arrows.meta}

\pdfoutput=1 
\hideLIPIcs  
\title{The TAG array of a multiple sequence alignment}
\author{Jannik Olbrich}{Ulm University, Germany}{jannik.olbrich@uni-ulm.de}{https://orcid.org/0000-0003-3291-7342}{}
\author{Enno Ohlebusch}{Ulm University, Germany}{enno.ohlebusch@uni-ulm.de}{https://orcid.org/0009-0008-3937-3652}{}
\authorrunning{J. Olbrich and E. Ohlebusch}
\Copyright{Jannik Olbrich and Enno Ohlebusch}



\category{}
\acknowledgements{We would like to thank Travis Gagie for suggesting this problem. Without him, this paper would not exist.}
\nolinenumbers 

\tikzset{gnode/.style={thick,draw,minimum size=0.6cm,circle,inner sep=0pt}}
\tikzset{gedge/.style={thick,draw=gray!80!black}}
\tikzset{gdedge/.style={thick,draw=gray!80!black,-{Stealth[length=2.5mm]}}}

\let\originalleft\left
\let\originalright\right
\renewcommand{\left}{\mathopen{}\mathclose\bgroup\originalleft}
\renewcommand{\right}{\aftergroup\egroup\originalright}

\newtheorem{fact}{Fact}

\newcommand{\GSA}[0]{\mathsf{GSA}}
\newcommand{\BWT}[0]{\mathsf{BWT}}
\newcommand{\EBWT}[0]{\mathsf{EBWT}}
\newcommand{\LF}[0]{\mathsf{LF}}
\newcommand{\TAG}[0]{\mathsf{TAG}}
\newcommand{\bra}[1]{\left(#1\right)}
\newcommand{\set}[1]{\left\{#1\right\}}
\newcommand{\ceil}[1]{\left\lceil#1\right\rceil}
\newcommand{\floor}[1]{\left\lfloor#1\right\rfloor}
\newcommand{\abs}[1]{\left|#1\right|}
\newcommand{\lexlt}[0]{<_\mathrm{lex}}

\newcommand{\intervCC}[2]{\left[#1\,..\,#2\right]}
\newcommand{\intervCO}[2]{\left[#1\,..\,#2\right)}
\newcommand{\intervOC}[2]{\left(#1\,..\,#2\right]}
\newcommand{\intervOO}[2]{\left(#1\,..\,#2\right)}
\newcommand{\bwtpos}[0]{\mathsf{bwt}_\mathsf{pos}}
\newcommand{\suff}[2]{\operatorname{suf}_{#2}\bra{#1}}
\newcommand{\alignment}[0]{\mathsf{alignment}}

\newcommand{\select}[1]{\mathit{select}_{#1}}
\newcommand{\rank}[1]{\mathit{rank}_{#1}}

\colorlet{factor_col}{brown!75!white!90!gray}

\def\alignvspace{0.36}
\def\alignhspace{0.27}
\newcommand{\adjustedexampleaccess}[2]{
	\edef\curchar{\csname example#1at#2\endcsname}%
	\IfStrEq{\curchar}{\%}{$\texttt{\textdollar}_\the\numexpr#1+1\relax$}{\curchar}%
}
\newcommand{\alignm}[3]{
	\StrLen{#2}[\strlen]%
	\xdef\alignmi{0}%
	\foreach \chr in {1,...,\strlen}{%
		\StrChar{#2}{\chr}[\chhhhh]%
		\if-\chhhhh%
			\relax%
		\else%
			\expandafter\xdef\csname example#1at\alignmi\endcsname{\chhhhh}%
			\xdef\chhhhh{\noexpand\adjustedexampleaccess{#1}{\alignmi}}%
			\xdef\alignmi{\the\numexpr\alignmi+1\relax}%
		\fi%
		\node (#1x\chr) at (\chr*\alignhspace,\alignvspace*#3) { \texttt{\strut\chhhhh} };
	}%
	\expandafter\xdef\csname example#1len\endcsname{\alignmi}%
}
\newcommand{\bwtchar}[2]{
	\ifnum#2=0%
		\adjustedexampleaccess{#1}{\the\numexpr\csname example#1len\endcsname-1\relax}%
	\else%
		\adjustedexampleaccess{#1}{\the\numexpr#2-1\relax}%
	\fi%
}
\newcommand{\examplerotation}[2]{
	\edef\curlen{\csname example#1len\endcsname}%
	\foreach \i in {0,...,\the\numexpr\curlen-1\relax} {%
		\edef\idx{\the\numexpr\i+#2\relax}%
		\ifnum\idx<\curlen%
			\relax%
		\else%
			\edef\idx{\the\numexpr\idx-\curlen\relax}%
		\fi%
		\adjustedexampleaccess{#1}{\idx}%
	}
}

\begin{document}

\maketitle

\begin{abstract}
Modern genomic analyses increasingly rely on pangenomes, that is, representations of the genome of entire populations.
The simplest representation of a pangenome is a set of individual genome sequences.
Compared to e.g.\ sequence graphs, this has the advantage that efficient exact search via indexes based on the Burrows-Wheeler Transform (BWT) is possible, that no chimeric sequences are created, and that the results are not influenced by heuristics.
However, such an index may report a match in thousands of positions even if these all correspond to the same locus, making downstream analysis unnecessarily expensive.
For sufficiently similar sequences (e.g. human chromosomes), a multiple sequence alignment (MSA) can be computed.
Since an MSA tends to group similar strings in the same columns, it is likely that a string occurring thousands of times in the pangenome can be described by very few columns in the MSA.
We describe a method to tag entries in the BWT with the corresponding column in the MSA and develop an index that can map matches in the BWT to columns in the MSA in time proportional to the output.
As a by-product, we can efficiently project a match to a designated reference genome, a capability that current pangenome aligners based on the BWT lack.
\end{abstract}

\section{Introduction}

Genomic analyses and diagnostics is often reference-based, that is, samples are compared with a reference genome of e.g.\ humans to speed up the analysis.
For a long time, such a reference consisted of a single genome.
However, a single reference sequence cannot capture the genetic diversity of a population.
As a consequence, modern tools focus on representations that include (common) variations.
Such a representation of the genomes of a population is commonly called a \emph{pangenome}.\footnote{
	The concept of a pangenome was originally developed for bacterial studies \cite{tettelin2005genome}, but now
	can refer to the entire genomic variation of any population.
	In this paper, we focus on the pangenome of a species.
}
A common use case for pangenome representations is \emph{read mapping}, where one seeks to determine the locus in the genome of a short substring of a DNA sample called a \emph{read}, while accounting for sequencing errors and natural variation.
Using pangenome representations can significantly reduce reference bias and thus mapping errors \cite{buechler2023efficient,garrison2018variation,kim2019graph,li2024bwt,siren2014indexing}.

Pangenome representations are mostly based on \emph{sequence graphs} (see e.g.\ \cite{andreace2023comparing,baaijens2022computational}). 
However, no widely agreed-upon metric for the quality of such graphs exists because several desirable qualities are at odds with each other \cite{baaijens2022computational}.
For example, using the same subgraph to represent a variant that occurs in multiple sequences may lead to a smaller graph, but may also induce chimeric sequences (i.e., sequences that are valid in the graph but do not exist in the pangenome), e.g.\ by allowing the combination of variants that do not occur together in nature.
Additionally, under the strong exponential-time hypothesis (SETH) it is impossible to index a sequence graph in polynomial time such that string matching queries can be answered in sub-quadratic time \cite{equi2023graphs}. For this reason, methods based on sequence graphs either resort to heuristic matching or cannot guarantee a good worst-case time complexity.
In contrast, there are indexes for strings or sets of strings which achieve optimal construction and query time complexities \cite{ferragina2000opportunistic}.
While these data structures would require space proportional to the combined size of all haplotypes present in a sequence graph and thus be far less practical,
almost as good time complexities can be achieved with indexes based on the run-length compressed Burrows-Wheeler Transform (BWT) \cite{bwt} such as variants of the $r$-index \cite{gagie2018optimal} (see e.g.\ \cite{bannai2020refining,bertram2024mover,cobas2025fast,nishimoto2021optimal}).

In combination with algorithms to construct these indexes for such huge datasets \cite{boucher2019prefix,diazdominguez2022grlbwt,li2024bwt,masillo2023cmsbwt,olbrich2025bwt,oliva2023recursive},
this opens the possibility to represent a population just by the set of sequences and not worry about e.g.\ graph indexing.

One problem of such a simple representation is that now a match may occur at thousands of positions in the index even when those occurrences all correspond to the same locus.
However, in practice, it is often desirable to locate a match relative to a linear reference sequence.
Recent works that use the BWT for pangenome indexing are for instance \emph{ropebwt3} \cite{li2024bwt} and \emph{Moni-align} \cite{varki2025accurate}.
The former aligns samples using the BWT-based index, while the latter merely uses the index to search for seeds (maximal exact matches between the pangenome and the pattern) and then aligns only to the most promising sequences in the pangenome.


 Recently, we showed how to compute a \emph{multiple sequence alignment (MSA)} of even very long sequences (such as human chromosomes), given that those sequences are very similar \cite{olbrich2025alignments}.
In an MSA, equal substrings that correspond the same locus are aligned. Therefore, we can identify the aforementioned thousands of positions in the index by just the corresponding column in the MSA.
Note that it is trivial to directly map such a column in the MSA to a single reference sequence if that reference is part of the alignment.
Tools such as ropebwt3 ``cannot project the alignment [of a pattern] to a designated reference genome'' \cite{li2024bwt}, so
``it seems [..] interesting to know which column of the alignment a character in the BWT comes from'' \cite{gagie2024tag}.

\subsection{Our contributions}
In this paper, we describe a space-efficient index that reports the columns of the alignment where a string occurs.
In particular, we build an index that is able to quickly report the distinct columns in the MSA where a match occurs.
To this end, we use the TAG array \cite{balaz2024wheelermaps,gagie2024tag}, which lists the columns in the alignment for each suffix in lexicographic order.
We first show how the run-length encoded TAG array can be built in linear time and small space,
and describe a sampled index and use known document-listing techniques to enable reporting of the distinct TAG values in a BWT interval in optimal time.
As an example application, we demonstrate that we can map multiple exact matches (MEMs) in the BWT-based index of ropebwt3 to a linear reference sequence quickly while using small space.

We focus on pangenomes for very closely related genomes, e.g.\ those of humans, since an MSA cannot sensibly represent the variations otherwise.  

\subsection{Related work}
A TAG array associates some data (a ``tag'') with each BWT position. In our case, this tag is a column in the MSA, but there are other tags that result in a run-length compressible TAG array:
In \cite{eskandar2025lossless}, positions in the BWT are tagged with the corresponding coordinates in a sequence graph.
Compared to our algorithms, theirs are more complex and also orders of magnitude more expensive to run in terms time and memory consumed, but it should be noted that it may arguably be the case that tagging BWT positions with graph positions is inherently harder than tagging them with columns of an MSA.
In \cite{depuydt2025run}, positions in the BWT are tagged with metagenomic class identifiers, resulting in an index that can be used to perform accurate metagenomic read classification in small space.

The remainder of this paper is structured as follows: \Cref{sec:preliminiaries} introduces the definitions and notations used throughout this paper.
\Cref{sec:tag_construction} describes our algorithm for constructing the TAG array, and in \Cref{sec:sampled} we describe a strategy and corresponding index
for sampling only a fraction $\frac{1}s$ of the TAG runs.
In \Cref{sec:experiments} we experimentally evaluate our algorithms before \Cref{sec:conclusion} concludes the paper.

\section{Preliminaries}
\label{sec:preliminiaries}

For $i, j\in\mathbb N_0$ we denote the set $\set{k\in\mathbb N_0 \mid i\leq k\leq j}$ by the interval notations $\intervCC{i}{j} = \intervCO{i}{j+1} = \intervOC{i-1}{j} = \intervOO{i-1}{j+1}$.
A \emph{string} (or \emph{array}) $S$ of \emph{length} $n$ over an \emph{alphabet} $\Sigma$ is a sequence of $n$ characters from $\Sigma$. We denote the length $n$ of $S$ by $\abs S$ and the $i$th symbol of $S$ by $S[i-1]$, i.e., strings and arrays are zero-indexed.
The \emph{substring} (or \emph{subarray}) of $S$ from $i$ to $j$ is denoted by $S\intervCC{i}{j} = S\intervCO{i}{j+1} = S\intervOC{i-1}{j} = S\intervOO{i-1}{j+1} = S[i]S[i+1]\dots S[j]$. For $i > j$, $S\intervCC{i}{j}$ is the \emph{empty string} $\epsilon$ of length $0$.
A substring of the form $S\intervCO{i}{n}$ is a \emph{suffix} of $S$ and is denoted by $\suff{S}{i}$.
A \emph{bit vector} is an array over the binary alphabet $\set{0,1}$.

\begin{figure}
	\centering
	\begin{tikzpicture}
\begin{scope}[yshift=-3cm*\alignvspace,xshift=-4cm]
	\alignm{0}{ACG-ACT\%}{3}
	\alignm{1}{AA--ACT\%}{2}
	\alignm{2}{ACGCAGT\%}{1}
	\path let \p1 = (0x1.west) in node[anchor=west] at (\x1,4.5*\alignvspace) {MSA};
\end{scope}

\begin{scope}[xshift=0cm]
\node[anchor=east] at (-0.7,1.5*\alignvspace) {$i$};
\node[anchor=east] at (2.3,1.5*\alignvspace) {$\TAG$};
\node[anchor=west] at (-0.5,1.5*\alignvspace) {$\BWT$};
\newcommand{\vbwt}[4]{
	\node[anchor=east] at (-0.7,-#1*\alignvspace) {$#1$};
	\node[anchor=east] (tagx#1) at (2.3,-#1*\alignvspace) {$#2$};
	\node[anchor=west] at (-0.5,-#1*\alignvspace) {\texttt{\bwtchar{#3}{#4}}};%
	\expandafter\xdef\csname exampletagval#1\endcsname{#2}%
	\expandafter\xdef\csname exampleisa#3at#1\endcsname{#4}%
	\xdef\exampleN{#1}%
	\def\curcolor{%
		\ifnum#4=0%
			factor_col%
		\else%
			black%
		\fi%
	}%
	\node[anchor=west] at (0,-#1*\alignvspace) {\textcolor{\curcolor}{\texttt{\examplerotation{#3}{#4}}}};%
}
\vbwt{0}{7}{0}{6}
\vbwt{1}{7}{1}{5}
\vbwt{2}{7}{2}{7}
\vbwt{3}{0}{1}{0}
\vbwt{4}{1}{1}{1}
\vbwt{5}{0}{0}{0}
\vbwt{6}{0}{2}{0}
\vbwt{7}{4}{0}{3}
\vbwt{8}{4}{1}{2}
\vbwt{9}{4}{2}{4}
\vbwt{10}{3}{2}{3}
\vbwt{11}{1}{0}{1}
\vbwt{12}{1}{2}{1}
\vbwt{13}{5}{0}{4}
\vbwt{14}{5}{1}{3}
\vbwt{15}{2}{0}{2}
\vbwt{16}{2}{2}{2}
\vbwt{17}{5}{2}{5}
\vbwt{18}{6}{0}{5}
\vbwt{19}{6}{1}{4}
\vbwt{20}{6}{2}{6}
\end{scope}

\begin{scope}[xshift=5cm,yshift=\alignvspace*-1cm]
	\xdef\oldalignhspace{\alignhspace}
	\def\alignhspace{0.5}
	\alignm{0}{ACG-ACT\%}{1}
	\foreach \c/\nexttag [count=\i from 0] in {11/1,15/2,7/4,-1/-1,13/5,18/6,0/7,5/0} {
		\node (tag\i) at ({(\i+1)*\alignhspace},-1*\alignvspace) {\strut\footnotesize\i};
		\ifthenelse{\equal{\c}{-1}}{}{
			\node (ibwt\i) at ({(\i+1)*\alignhspace},0*\alignvspace) {\strut\footnotesize\c};
			\node (nexttag\i) at ({(\i+1)*\alignhspace},-2*\alignvspace) {\strut\footnotesize\nexttag};

			\draw[->] (nexttag\i) -- ++(0,-2) node[midway,sloped,above,yshift=-0.1cm,anchor=south west,xshift=-0.9cm]{\footnotesize$\TAG[\c]=\nexttag$} to[out=-90,in=0] (0cm,-3cm-\i*0.4cm) to[out=180,in=0] (tagx\c);
		}
	}
	\node[left=0.1cm of tag0] {column};
	\node[left=0.1cm of ibwt0] {$\BWT^{-1}$};
	\node[left=0.1cm of nexttag0] {$\mathit{tag}$};
	\xdef\alignhspace{\oldalignhspace}
\end{scope}
\end{tikzpicture}
	\caption{
	An example MSA (top left) with the corresponding $\BWT$ (centre, instead of the sorted suffixes we display the sorted rotations for clarity).
	Strings represented in the MSA are coloured ({\color{factor_col}$\blacksquare$}).
	On the right, the first string is displayed with the corresponding positions in the $\BWT$ and TAG values.
	}
	\label{fig:tag_example}
\end{figure}
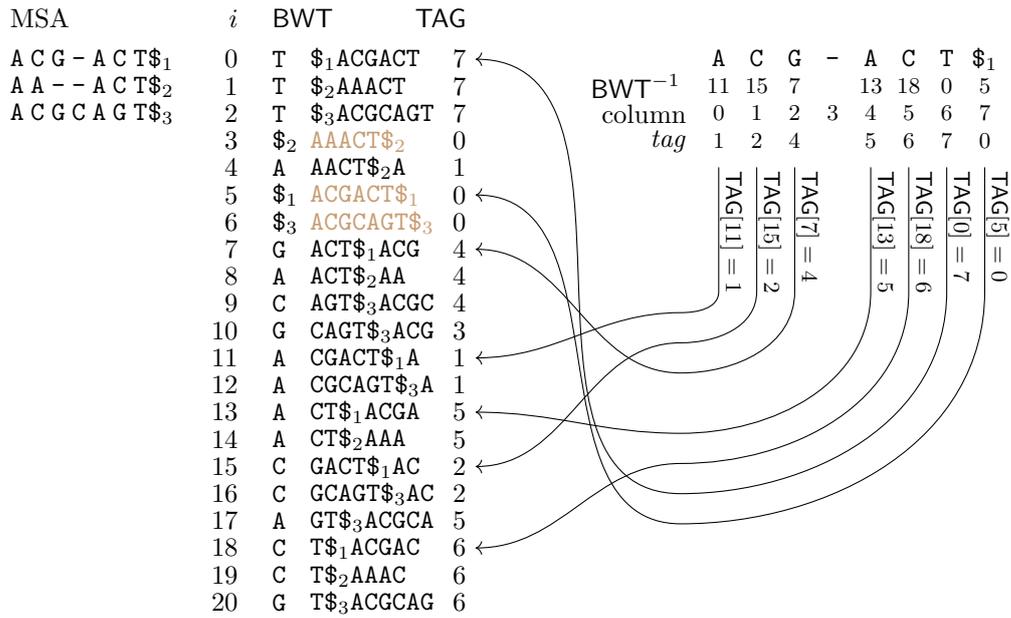

A \emph{multiple sequence alignment} (MSA) of a set of strings $\mathcal S$ is obtained by inserting a number of \emph{gap characters} `\texttt{-}' into each string in $\mathcal S$ such that the resulting strings all have the same length.
An example MSA of the strings $\set{\texttt{ACGACT\$}, \texttt{AAACT\$}, \texttt{ACGCAGT\$}}$ is given in the top-left of \Cref{fig:tag_example}.
A ``good'' MSA has few gap-symbols and columns mostly contain the same character.
In this paper, we are not concerned with the precise optimization objective or methods for construction, so we refer the interested reader to \cite{chatzou2016multiple} for an overview.
However, it is noteworthy that any ``good'' MSA of sufficiently similar sequences will exhibit \emph{contextual locality}, that is, the suffixes starting in a column in the MSA are likely to be similar (i.e., share a long prefix) \cite{balaz2024wheelermaps,gagie2024tag}.
For instance, the suffixes starting in the fifth column of the MSA in \Cref{fig:tag_example} are $\texttt{ACT\$}$, $\texttt{ACT\$}$ and $\texttt{AGT\$}$.

We assume totally ordered alphabets. This induces a total order on strings.
Specifically, we say a string $S$ of length $n$ is \emph{lexicographically smaller} than a string $T$ of length $m$ if and only if there is some $\ell<\min\set{n,m}$ such that $S\intervCO{0}{\ell} < T\intervCO{0}{\ell}$ and either $n = \ell < m$ ($S$ is a prefix of $T$) or $\ell<\min\set{n,m}$ and $S[\ell] < T[\ell]$, and write $S\lexlt T$ in this case.

Given a string $S$, $\rank{c}\bra{S, i}$ is the number of $c$'s occurring in $S$ up to (but excluding) index $i$, i.e., $\rank{c}\bra{S, i} = \abs{\set{j\in\intervCO0i\mid S[j]=c}}$.
The \emph{select} function returns the index of the $i$th occurrence of $c$ (zero-based) in $S$, i.e., $\rank{c}\bra{S, \select{c}\bra{S, i}}=i$.

The \emph{generalized suffix array} for a collection of strings $S_1,\dots,S_n$ is an array $\GSA$ where $\GSA[i] = (k,j)$ indicates that there are $i$ lexicographically smaller suffixes than $\suff{S_k}{j}$ among the suffixes of $S_1,\dots,S_n$.
We assume the strings to be \emph{dollar-terminated}, that is, the last character of each string $S_i$ is `$\texttt{\$}$', which is smaller than all other characters.
In the case of a tie between equal suffixes, we define the one occurring earlier in the input to be smaller.\footnote{This is equivalent to using $n$ terminal symbols $\texttt{\$}_1<\dots<\texttt{\$}_n$ and terminating $S_i$ with $\texttt{\$}_i$. Hence the name `multidollar-$\EBWT$.'}

Throughout this paper, we use the multidollar-$\EBWT$ as the BWT for string collections (see \cite{cenzato2024survey} for an overview of such BWT variants).
In the remainder of this paper, we refer to the multidollar-$\EBWT$ just by $\BWT$.
It is defined as follows. Let $\GSA[i] = (k,j)$. Then $\BWT[i] = S_k[j-1]$ if $j>0$ and $\BWT[i] = S_k[\abs{S_k}-1] = \text{`$\texttt{\$}_k$'}$ otherwise.
An example can be seen in Figure~\ref{fig:tag_example}.
The \emph{$\LF$-mapping} is a function such that $\LF[i] = \GSA^{-1}[(k,j-1)]$ for $\GSA[i]=(k,j)$ if $j > 0$ and $\LF[i] = \GSA^{-1}[(k,\abs{S_k}-1)]$ otherwise.
$\LF$ can thus be used to iterate over an input string in reverse order, given the index $p$ of the last character of the input string in $\BWT$.
Specifically, $S_k = \BWT[\LF^{\abs{S_k}-1}[p]]\dots\BWT[\LF^1[p]]\BWT[\LF^0[p]]$ where $\GSA[p] = (k,0)$.

Although we use the dollar-$\EBWT$, the algorithms presented in this paper are applicable to all BWT variants where for each input string $S$ there is an index $i$ such that the indices $\LF^0[i],\LF^1[i],\dots,\LF^{\abs S-1}[i]$ are all distinct and $S=\BWT[\LF^{\abs S-1}[i]]\dots \BWT[\LF^0[i]]$.\footnote{
	Notably, this excludes the original EBWT for non-primitive input strings, because there each root of such a string corresponds to a distinct cycle in $\LF$.
	As far as we can tell, all other BWT variants for string collections satisfy this requirement.
}

\begin{definition}[$\TAG$ \cite{gagie2024tag}]
	Let $\GSA[i] = (k,j)$. Then $\TAG[i]$ is the column in the alignment of character $j$ of string $k$.
	\label{def:tag}
\end{definition}
Note that this definition gives an immediate linear-time algorithm for computing $\TAG$:
For each character in the alignment, use the inverse of $\GSA$ to find the corresponding position in $\TAG$ and write the character's column.
However, this na{\"i}ve approach requires holding the inverse of $\GSA$ in main memory.
Definition~\ref{def:tag} is equivalent to the following definition based on the $\BWT$ instead of the suffix array, which follows immediately from the definition of the $\BWT$ given above.
\begin{fact}
	$\TAG[i]$ is the column in the alignment of the character immediately following $\BWT[i]$ in the dataset.
	\label{fact:tag_via_bwt}
\end{fact}
We will use this definition from now on as it does not depend on $\GSA$.
In simpler terms, consider a character in column $i$ in row $s$ which corresponds to $\BWT[j]$.
Then $\TAG[j]$ is $\mathit{col}$, where $\mathit{col}$ is the first column after $i$ where there is a non-gap character in row $s$.
An example of the $\TAG$ array of a multiple string alignment (MSA) is shown in Figure~\ref{fig:tag_example}.
For instance, the base $\texttt{G}$ in the first string corresponds to $\BWT[9]=\texttt{G}$, and the character following this $\texttt{G}$ in the alignment occurs in column $4$.
Therefore, we have $\TAG[9] = 4$.

In the next section, we present an algorithm that requires only access to the $\BWT$, $\LF$ and the positions of the gaps in the alignment to compute the run-length encoded TAG array.

\section{Computing the TAG array}
\label{sec:tag_construction}

Before describing our algorithm we briefly recall how we can reconstruct an input string from the $\BWT$ using $\LF$.
For this, we start at the position $p$ of the string's last character (a `$\texttt\$$' in our case) in the $\BWT$.\footnote{Equivalently, the rank of the input string in the sorted list of all rotations. Many tools for computing the $\BWT$ directly output these indices \cite{boucher2019prefix,olbrich2025bwt,oliva2023recursive}.}
$\LF[p]$ now gives us the position in the $\BWT$ of the character preceding the `$\texttt\$$' in the string, $\LF[\LF[p]]$ gives us the character preceding that, and so on.

Essentially, we compute $\TAG$ using the inverse of Fact~\ref{fact:tag_via_bwt}:
for each TAG value $\mathit{tag}$, we find the indices where the TAG array $\TAG$ has value $\mathit{tag}$.
Consider a column $\mathit{col}$.
By Fact~\ref{fact:tag_via_bwt}, $\TAG$ should have value $\mathit{col}$ at the indices where
the characters preceding the characters in column $\mathit{col}$ are in $\BWT$.

We consider each column of the alignment from right to left and maintain the BWT positions by iterating over the input strings from right to left using $\LF$ as is done when reconstructing the strings from $\BWT$.
By incorporating the positions of the gap symbols, this is synchronized such that we consider a column of the alignment at each time step.

Note that, given the indices of the characters in column $\mathit{col}$ in the $\BWT$,
the indices of the preceding characters can be found with a single $\LF$ step each.
If a symbol in the alignment in column $\mathit{col}$ is a gap (\texttt{`-'}), the first preceding non-gap character has a TAG value greater than $\mathit{col}$.
Therefore, in this case we just ignore the current column in this string.

\begin{algorithm}
	\For{$i=0\to n-1$}{
		$\bwtpos[i] = \LF\bra{p_i}$ where $p_i$ is the position of the $\texttt\$$ of string $i$ in $\BWT$\;
	}
	\For{$\mathit{col} = m - 1\to 0$}{
		\For{$i=0\to n-1$}{
			\If{$\alignment[i][\mathit{col}] \neq \texttt{`-'}$}{
				$\TAG[\bwtpos[i]] \gets \mathit{col}$\;
				$\bwtpos[i]\gets\LF\bra{\bwtpos[i]}$\;
			}
		}
	}
	\caption{
	Simple algorithm for computing $\TAG$ from an alignment of length $m$ of $n$ strings.
	Throughout the algorithm, $\bwtpos[i]$ is the position in the $\BWT$ of the character preceding the column $\mathit{col}$ in string $i$.
	}
	\label{alg:simple_tag}
\end{algorithm}

Algorithm~\ref{alg:simple_tag} shows the procedure.
Note that any two iterations of the second for-loop of Algorithm~\ref{alg:simple_tag} concern two different $\mathit{col}$ values (columns).
Therefore, no two iterations of this loop can concern the same TAG run.
It is thus possible to immediately compute the TAG runs without having to store the entire $\TAG$ array.

Modifying Algorithm~\ref{alg:simple_tag} accordingly, we obtain an algorithm that outputs the TAG runs in order of decreasing TAG value.
To obtain the run-length compressed TAG array, it is hence necessary to sort the TAG runs by (start or end) index afterwards.




\subsection{Practical construction}
In the previous section, it is noted that one can use Algorithm~\ref{alg:simple_tag} to directly compute the TAG runs.
Indeed, in the inner for-loop, all indices $i$ in $\TAG$ are considered where $\TAG[i]=\mathit{col}$ for a specific value of $\mathit{col}$.
Thus, it is possible to collect all these indices and then sort them in increasing order after the inner for-loop.
Each run of consecutive indices in the resulting sorted list then corresponds to a TAG run of value $\mathit{col}$.
In the following, we say that there is a TAG run of value $\mathit{tag}$ from $l$ to $r$ ($l\leq r$) if $\TAG[l]=\dots=\TAG[r]=\mathit{tag}$ and denote this with $\mathit{tag}\text-[l,r]$.

The number of objects involved in sorting and the number of $\LF$ accesses of the approach just described can be reduced with the following observations:
\begin{itemize}
	\item If $\TAG[i] = \TAG[j]$, it is likely that $\TAG[\LF[i]] = \TAG[\LF[j]]$ \cite{gagie2024tag}.
	\item If $i$ and $j$ belong to the same $\BWT$ run, we have $\LF[j] = \LF[i] + (j - i)$ \cite{ferragina2000opportunistic}.
\end{itemize}
Because both the TAG array and the $\BWT$ have contextual locality \cite{balaz2024wheelermaps,gagie2024tag}, two indices in the same TAG run likely belong to the same $\BWT$ run.
Therefore, a TAG run $\mathit{tag}\text-[l,r]$ likely implies a TAG run $(\mathit{tag}-1)\text-[\LF[l],\LF[r]]$.
We can thus often operate on these runs instead of on the individual strings.

We maintain a set of TAG runs such that after each iteration of the inner for-loop, these TAG runs are maximal and disjoint.
Each such TAG run is associated with the set of strings corresponding to the contained BWT indices.
For each $\mathit{col}$ we thus
\begin{enumerate}
	\item remove indices from their current TAG run where the corresponding row in the MSA has a gap symbol in the current column,
	\item insert indices where the corresponding row in the MSA had a gap symbol in the column processed in the previous iteration (but a base in the current column),
	\item perform the $\LF$ step for each run, and
	\item merge adjacent runs (e.g.\ applying $\LF$ to the TAG runs $5-[13,14]$ and $5-[17,17]$ in \Cref{fig:tag_example} results in $4-[7,8]$ and $4-[9,9]$, which can clearly be merged) and output the result.
\end{enumerate}
Note that, during the $\LF$ step it may be possible that a run is split if it crosses a $\BWT$ run boundary.
In our example, this happens with the TAG run $6-[18,20]$ which results in the TAG runs $5-[13,14]$ and $5-[17,17]$.
Thus, the data structure used for maintaining the indices in a run must support concatenation (for merging), splitting, and removal of an element (which may result in two distinct runs).
Data structures which support these operations in (amortized) $\mathcal O(\log n)$ time are e.g.\ Red-Black trees or Splay trees \cite{sleator1985self,tarjan1983data}.
We use Splay trees because of their simpler implementation.

\section{A sampled $\TAG$ array}
\label{sec:sampled}
Because both the MSA and BWT possess contextual locality, the TAG array of an MSA of similar strings is likely run-length compressible \cite{gagie2024tag}.
One can facilitate random access to the run-length compressed TAG array using a sparse bit vector indicating the boundaries of the TAG runs.
Assuming $r_\TAG$ TAG runs and an alignment with length $m$ and $N$ non-gap characters,
this would require at least $r_\TAG \cdot \log_2 m + r_\TAG \cdot \log_2\frac{N}{r_\TAG}$ bits for the TAG run labels and the run boundaries.

However, $r_\TAG$ is likely to be larger than the number $r$ of runs of the $\BWT$ (and guaranteed to be at least the length $m$ of the alignment).
Additionally, the alphabet used for $\TAG$ is much larger than that for $\BWT$.
Therefore, a na{\"i}ve run-length compressed TAG array would require many times as much memory as the run-length compressed $\BWT$, with the majority of the memory used for the run labels.
In this section we present a technique for sampling the TAG array where we store the labels of only a fraction $\frac{1}{s}$ of the TAG runs, for some user-defined parameter $s$.
For this, $\LF$ is assumed to be available.\footnote{This is not a major restriction because efficient access to $\LF$ can be facilitated using space proportional to the number of runs $r$ of the $\BWT$ \cite{gagie2018optimal}, and $\LF$ is available anyway in many tools using the $\BWT$.}

Note that there is no obvious method to just sample every $s$-th TAG value and use $\LF$ to walk to the next sampled TAG run for a query.
This is because, regardless of which cells are chosen as the starting points for $\LF$ steps in the TAG runs, the $\LF$ steps may ``jump over'' sampled TAG runs.\footnote{
	In any sensible MSA, there is no column which contains only gap symbols.
	Note however, that there may very well be TAG runs where each corresponding symbol in the alignment is preceded by a gap symbol.
}
As a consequence, it would not be possible to guarantee that a sampled TAG run can be reached with a given number of $\LF$ steps. 
In the following, we present a more involved sampling strategy which guarantees that we can always reach a sampled TAG run with fewer than $s$ $\LF$ steps.

\begin{figure}[ht]
	\centering
	\begin{tikzpicture}
	\def\hspace{0.6}
	\begin{scope}
		\foreach \l [count=\i from 0] in {18,19,20,1,3,0,2,15,4,10,16,5,6,7,8,11,12,9,13,14,17} {
			\expandafter\xdef\csname examplelf\i\endcsname{\l}
		}
		\xdef\lastseentag{-1}
		\newcounter{examplenumtagruns}
		\setcounter{examplenumtagruns}{-1}
		\foreach \i in {0,...,\exampleN} {
			\edef\curtag{\csname exampletagval\i\endcsname}
			\ifthenelse{\equal{\lastseentag}{\curtag}}{}{
				\addtocounter{examplenumtagruns}{1}
				\expandafter\xdef\csname exampletagboundary\theexamplenumtagruns\endcsname{\i}
			}
			\expandafter\xdef\csname exampletagtorleidx\i\endcsname{\theexamplenumtagruns}
			\xdef\lastseentag{\curtag}
		}
		\foreach \i in {0,...,\exampleN} {
			\node[anchor=east] (i\i) at (\i*\hspace,\alignvspace) {\strut $\i$};
			\edef\curtag{\csname exampletagval\i\endcsname}
			\edef\curtagrun{\csname exampletagtorleidx\i\endcsname}
			\edef\curboundary{\csname exampletagboundary\curtagrun\endcsname}
			\node[anchor=east] (lf\i) at (\i*\hspace,-2*\alignvspace) {\strut $\ifthenelse{\equal{\curboundary}{\i}}{}{\textcolor{gray!50}}{\csname examplelf\i\endcsname}$};
			\node[anchor=east] (tag\i) at (\i*\hspace,0*\alignvspace) {\strut $\curtag$};
			\node[anchor=east] (rb\i) at (\i*\hspace,-1*\alignvspace) {\strut $\ifthenelse{\equal{\curboundary}{\i}}{1}{0}$};
		}
		\path let \p1 = (i0) in node[anchor=west] at (-1.5cm,\y1) {$i$};
		\path let \p1 = (tag0) in node[anchor=west] at (-1.5cm,\y1) {$\TAG$};
		\path let \p1 = (rb0) in node[anchor=west] at (-1.5cm,\y1) {$\mathbb{RB}$};
		\path let \p1 = (lf0) in node[anchor=west] at (-1.5cm,\y1) {$\LF$};
		\def\hspace{0.725}
		\begin{scope}[yshift=-4.5cm,xshift=-0.9mm]
		\xdef\dpos{0}%
		\foreach \i in {0,...,\theexamplenumtagruns} {
			\edef\j{\csname exampletagboundary\i\endcsname}%
			\edef\tagval{\csname exampletagval\j\endcsname}%
			\edef\targetrun{\csname exampletagtorleidx\csname examplelf\j\endcsname\endcsname}%
			\edef\targetj{\csname exampletagboundary\targetrun\endcsname}%
			\edef\targettagval{\csname exampletagval\targetj\endcsname}%
			\node[anchor=east] (e\i) at (\i*\hspace,-2*\alignvspace) {\strut $\targetrun$};
			\node[anchor=east] (tagp\i) at (\i*\hspace,-1*\alignvspace) {\strut $\tagval$};
			\node[anchor=east] (ii\i) at (\i*\hspace,0*\alignvspace) {\strut $\i$};
			\draw[->,gdedge] ([yshift=1.5mm]lf\j.south) --node[midway,above,sloped,yshift=-1mm]{\footnotesize $\rank{1}\bra{\mathbb{RB}, \csname examplelf\j\endcsname\!+\!1}\!-\!1\!=\!\targetrun$} ([yshift=-1mm]ii\i.north);
			\node[anchor=east] (d\dpos) at (\dpos*\hspace,-3*\alignvspace) {\strut $1$};
			\xdef\dpos{\the\numexpr\dpos+1\relax}%
			\ifnum\targettagval<\tagval\relax%
				\expandafter\ifnum\the\numexpr\targettagval+1\relax<\tagval\relax%
					\foreach \k in {\the\numexpr\targettagval+2\relax,...,\tagval} {
						\node[anchor=east] (d\dpos) at (\dpos*\hspace,-3*\alignvspace) {\strut $0$};
						\xdef\dpos{\the\numexpr\dpos+1\relax}%
					}
				\else%
					\xdef\dpos{\the\numexpr\dpos+\tagval-\targettagval-1\relax}%
				\fi%
			\fi%
		}
		\node[anchor=east] (d\dpos) at (\dpos*\hspace,-3*\alignvspace) {\strut $1$};
		\path let \p1 = (tagp0) in node[anchor=west] at (-1.5cm,\y1) {$\TAG'$};
		\path let \p1 = (e0) in node[anchor=west] at (-1.5cm,\y1) {$e$};
		\path let \p1 = (ii0) in node[anchor=west] at (-1.5cm,\y1) {$i$};
		\path let \p1 = (d0) in node[anchor=west] at (-1.5cm,\y1) {$\mathbb D$};
		\end{scope}
	\end{scope}
	\begin{scope}[xshift=10.5cm,yshift=-5cm]
		\def\radius{1.3cm}
		\foreach \i [count=\j from 0] in {7,4,8,6,3,0,10} { 
			\node[gnode] (n\i) at ($(\j*360/7:\radius)$) {$\i$};
		}
		\foreach \i [count=\j from 1] in {1,2} { 
		\node[gnode] (n\i) at ($(n0)-(\j*\radius-0.134*\radius,0.5*\radius)$) {$\i$};
		}
		\node[gnode] (n9) at ($(n4)+(0,\radius)$) {$9$};
		\node[gnode] (n5) at ($(n8)+(0,\radius)$) {$5$};

		\foreach \i in {0,...,\theexamplenumtagruns} {
			\edef\j{\csname exampletagboundary\i\endcsname}%
			\edef\tagval{\csname exampletagval\j\endcsname}%
			\edef\target{\csname exampletagtorleidx\csname examplelf\j\endcsname\endcsname}%
			\edef\targetj{\csname exampletagboundary\target\endcsname}%
			\edef\targettagval{\csname exampletagval\targetj\endcsname}%
			\ifnum\targettagval<\tagval%
				\draw[gdedge] (n\i) --node[midway,sloped,above]{\footnotesize$\the\numexpr\tagval-\targettagval\relax$} (n\target);
			\else%
				\draw[gdedge,draw=gray!50] (n\i) --node[midway,sloped,auto]{\footnotesize$\the\numexpr\tagval-\targettagval\relax$} (n\target) node[draw,strike out,pos=0.5,thick,red,sloped]{};
			\fi
		}
	\end{scope}
\end{tikzpicture}
	\caption{
	Top/Left: The TAG array, bit vector $\mathbb{RB}$ and $\LF$ for our example (cf.\ Figure~\ref{fig:tag_example}).
	Below, the run-length compressed TAG array $\TAG'$ is shown with the array $e$ derived from $\LF$.
	Values of $\LF$ that are not used (i.e., do not occur at the start of a TAG run) are printed in light gray.
	\newline
	Bottom right:
	The graph $G$ defined by $e$.
	We have $R=\set{1,3}$ because $\TAG'[e[1]]=\TAG'[e[3]]=\TAG'[0]=7\not<0=\TAG'[1]=\TAG'[3]$.
	For all other $i\in\set{0,\dots,10}$ we have $\TAG'[e[i]] < \TAG'[i]$.
	Above each edge $(i,e[i])$, the difference $\TAG'[e[i]] - \TAG'[i]$ is displayed.
	Edges from nodes in $R$ (i.e., where this difference is negative) are crossed out.
	Note that the remaining edges form a rooted forest where the roots are exactly the nodes in $R$.
	}
	\label{fig:sampling}
\end{figure}
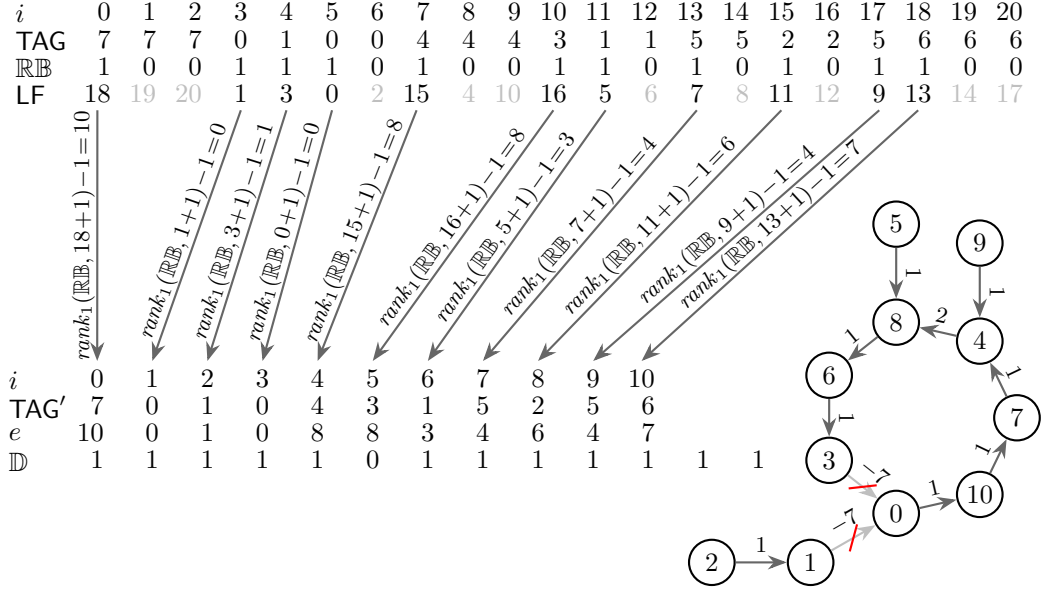
The following concepts and data structures are illustrated in Figure~\ref{fig:sampling} for our running example.

Let $e[i]$ be the index of the TAG run containing $\LF[b_i]$, where $b_i$ is the start of TAG run $i$.
Let $\mathbb{RB}$ be a (sparse) bit vector indicating the TAG run boundaries, i.e., $\mathbb{RB}[i] = 1$ if and only if $i$ is the start of a TAG run. 
We have $b_i = \select{1}\bra{\mathbb{RB},i}$ and $e[i] = \rank{1}\bra{\mathbb{RB}, \LF[b_i] + 1} - 1$.
Therefore, $e[i]$ can be computed with one $\LF$ computation and a $\select1$ and a $\rank1$ query and we do not have to store $e$ explicitly.

Now let $\TAG'$ be such that $\TAG'[i]$ is the TAG value of the $i$th TAG run.
Note that there are at most $n$ indices $i$ where $\TAG'[e[i]] \geq \TAG'[i]$, because the run head of such a TAG run must correspond to the first character in a string of the alignment. 
Let the set of these indices be $R$.
Now consider the digraph $G=(V,E)$ with $V=\set{0,\dots,r_\TAG-1}$ and $E = \{(i,e[i]) \mid i\in[0..r_\TAG-1]\setminus R\})$.
That is, $G$ is a graph where each TAG run is a node and there is an edge from a node $u$ to $v$ if and only if an $\LF$ step from the run-head of TAG run $u$ results in an index in TAG run $v$.
By definition, for each edge $(i,e[i]) \in E$ we have $\TAG'[e[i]] < \TAG'[i]$. Therefore, $G$ is acyclic.
Additionally, the out-degree of each node is at most $1$.
Thus, $G$ is not only a DAG but a rooted forest with $R$ as the roots and each edge pointing ``upwards'' towards a root (cf.\ Figure~\ref{fig:sampling}).

We require a set of sampled TAG runs to satisfy one constraint: Given a sampling rate $s$,
the distance (in $G$) from an unsampled node to a sampled node should be less than $s$.
Note that this implies that the roots $R$ of the forest are sampled.

Sampled TAG runs are marked in a bit vector $\mathbb{S}$.
Rank support on this bit vector is used to access an array $\mathsf L$ storing the TAG values of the sampled TAG runs.
We can thus decide in constant time whether a TAG run is sampled and retrieve the TAG value if it is.

To retrieve the TAG value of an unsampled TAG run $i$, we compute $e[i]$ as described above, recursively find the TAG value of TAG run $e[i]$, and add the difference $\TAG'[i] - \TAG'[e[i]]$ to the result.
The distance-constraint ensures that we can get any TAG value using fewer than $s$ recursion steps.
For this to work, we encode the difference $\TAG'[i] - \TAG'[e[i]]$ for all unsampled $i$ with unary encoding in a bit vector $\mathbb D$. For the sampled TAG runs, we also encode the value $1$ for simplicity.
Specifically, if the $i$th one-bit in $S$ is at position $p$, the $i+1$th one-bit is at position $p + \bra{\TAG'[i] - \TAG'[e[i]]}$ if TAG run $i$ is sampled and at position $p+1$ otherwise.
Two select queries on this bit vector then suffice to extract this difference and thus to compute $\TAG'[i]$ given $\TAG'[e[i]]$.
(We also an additional $1$ for an easier implementation, see \Cref{alg:sampling_access}.)

\begin{algorithm}
	\SetKwFunction{proc}{getTAG}
	\SetKwProg{myproc}{Procedure}{}{}
	\myproc{\proc{$i$}}{
	\lIf(\tcp*[f]{TAG run $i$ is sampled}){$\mathbb{S}[i] = 1$}{\Return $\mathsf L[\rank1\bra{\mathbb S,i}]$}
	$\mathit{difference}\gets \select1\bra{\mathbb D, i + 1} - \select1\bra{\mathbb D,i}$\tcp*{Because an $\bra{r_\TAG+1}$th $1$ is appended to $\mathbb D$, $\select1\bra{\mathbb D, i + 1}$ is defined for all $i\in[0..r_\TAG)$}
	$\mathit{runStart}\gets \select1\bra{\mathbb{RB}, i}$\tcp*{called $b_i$ in the text}
	$\mathit{precBwtPos}\gets\LF[\mathit{runStart}]$\;
	$\mathit{precRun}\gets \rank1\bra{\mathit{predBwtPos} + 1} - 1$\tcp*{$\mathit{precRun} = e[i]$}
	\Return $\mathit{difference} + \text{\proc{$\mathit{precRun}$}}$\;
	}
	\caption{Computing the TAG value of a TAG run, given its index $i$.}
	\label{alg:sampling_access}
\end{algorithm}
Algorithm~\ref{alg:sampling_access} shows how the TAG access is performed.

\noindent\textbf{Space complexity}\newline
Let $N$ be the total number of characters in the alignment, $m$ the length of the alignment and $g$ the total number of gap symbols preceding the run-heads of the sampled $\TAG$ runs.
In a good MSA, the number of gap symbols is expected to be very small compared to the size of the data set.
Additionally, because the number of TAG runs is also expected to be small, $g$ is expected to include only every $N/r_\TAG$th gap symbol.
We assume that there are $\frac1sr_\TAG$ sampled TAG runs. In the next section, it is shown that this is possible.

In addition to access to $\LF$, we need
\begin{itemize}
	\item the array $\mathsf L$ containing the sampled TAG values ($r_\TAG\cdot \ceil{\log_2 m}/s$ bits),
	\item the bit vector $\mathbb{RB}$ with $\rank1$ and $\select1$ support,
	\item the bit vector $\mathbb{S}$ with $\rank1$ support ($r_\TAG (1+o(1))$ bits), and
	\item the bit vector $\mathbb{D}$ with $\select1$ support ($(r_\TAG + g)(1+o(1))$ bits).
\end{itemize}
The bit vector $\mathbb{RB}$ can be implemented e.g.\ with a plain bit vector for $\mathcal O(1)$ rank/select operations ($N(1+o(1))$ bits) \cite{gonzalez2005practical},
or using the Elias-Fano representation for non-decreasing sequences where rank and select take $\mathcal O\bra{\log\frac{N}{r_\TAG}}$ and $\mathcal O(1)$ time, respectively ($r_\TAG\bra{2+\log_2\frac{N}{r_\TAG}}$ bits) \cite{elias1974efficient,fano1971number}.

\noindent\textbf{Time complexity}\newline
We obtain a time complexity of $\mathcal{O}\bra{s\cdot \bra{t_r+t_s+t_\LF}}$, where $t_r$, $t_s$ and $t_\LF$ are the time complexities of rank and select queries to $\mathbb{RB}$ and computing $\LF[i]$, respectively.

\subsection{Selecting the sampled TAG runs}
For selecting the sampled TAG runs, we use an (optimal) greedy algorithm that minimizes the number of sampled TAG runs under the above constraints.
Recall that we want to choose a smallest set $S\subseteq V$ of nodes in a forest $G=(V,E)$ such that it requires fewer than $s$ steps to reach a node in $S$ from any node not in $S$, where each step must go ``upwards'' (i.e., towards the corresponding root).
Let this smallest distance of a node $u$ upwards to the closest node $v\in S$ be $a(u)$ and call $v$ the \emph{witness of $u$}. For each node $u$ in $S$ we have $a(u)=0$.
Since the trees of the forest are independent of each other, we assume that the graph is a tree in the following. 

Note that the root of the tree must always be in $S$, and that removing a subtree can never increase the size of the smallest solution.
Now consider a node $u$ with maximum depth $d(u)$ (i.e., distance to the root).
Regardless of which ancestor $v$ of $u$ satisfying $d(u) - d(v) < s$ is chosen,
by selection of $u$, $v$ is a valid witness for all descendants of $v$ (because their distance to $v$ is at most $d(u) - d(v) < s$).
Therefore, we may remove the subtree rooted at $v$ (including $v$) and return $v$ plus the solution of the remaining tree.
Finally, it is optimal to choose $v$ such that $d(u) - d(v) < s$ is maximal.
This is because choosing any lower node would eliminate a strict subset of the descendants of $v$, and can therefore not lead to a better solution.
We thus obtain the following algorithm: While the tree is non-empty, find a node $u$ with maximum depth $d(u)$, determine the highest ancestor $v$ of $u$ that satisfies $d(u) - d(v) < s$, output $v$, and remove the subtree rooted at $v$.

Note that, for all choices except the last one where the root of the tree is chosen, $v$ can always be chosen such that $d(u) - d(v) = s - 1$.
Therefore, for each node $v$ in $S$ besides the root, there are at least $s-1$ nodes not in $S$ for which $v$ is the witness.
In a tree with $n$ nodes we therefore have $\abs S \leq \floor{\frac{n}{s}} + 1$.

The above algorithm can be emulated simpler than described above as follows:
for a node $u$, let $d'(u)$ be the maximum number of nodes on a ``downwards'' path that does not include nodes in $S$.
For each $u$ in $S$ we have $d'(u)=0$.
For each $u$ not in $S$, $d'(u) = 1 + \max\bra{\{d'(v) \mid v\text{ is child of }u\}\cup\{0\}}$ holds.
By the observations above, each node in $S$ besides the root has a child $v$ with $d'(v)=s-1$. Conversely, every node with a child $v$ with $d'(v) \geq s-1$ must be in $S$.
Since $d'(u)$ depends only on $u$'s children, $d'$ (and thus $S$) can be computed during a bottom-up traversal.
This immediately gives a simple optimal bottom-up traversal algorithm for deciding which nodes are in $S$.

\noindent\textbf{Directly constructing the sampled index}\newline
Note that, during the construction algorithm presented in \Cref{sec:tag_construction}, we traverse the TAG runs in exactly such a bottom-up order as required for the optimal sampling algorithm above.
It is therefore possible to immediately select the sampled TAG runs during the construction.

For this, we use a dynamic bit vector to mark the TAG run heads.
The sampled TAG values are stored together with the start position of the respective TAG run, and then sorted afterwards according to this position.
Finally, for the unsampled TAG runs, we need to store the difference $\TAG'[i] - \TAG'[e[i]]$.
As explained in the previous section, the number of preceding gaps is usually small. Therefore, there are only few runs where $\TAG'[i] - \TAG'[e[i]]$ is greater than one.
We store only these differences in combination with the start position of the respective TAG run.
From these TAG run boundaries, sampled tag runs, and unsampled tag runs $i$ with $\TAG'[i] - \TAG'[e[i]] > 1$ we can then construct all data structures needed for the sampled TAG index.

\subsection{Reporting the distinct TAGs in a $\BWT$ interval}
\label{sec:distinct}
Given an interval $[l,r]$ (e.g.\ resulting from a backward search on the $\BWT$), we would like to report the set of (distinct) TAG values in $\TAG[l,r]$.
In particular, the time complexity of this operation should depend only on the size of the output and not on the size $r+1-l$ of the interval $[l,r]$.
Note that we can use $\rank1$ queries on $\mathbb{RB}$ to ``translate'' the interval $[l,r]$ such that it refers to the run-length compressed TAG array $\TAG'$ instead of $\TAG$ (without affecting the set of distinct TAG values).
We therefore consider the problem of reporting the distinct TAG values in $\TAG'[l,r]$.


To achieve this, we use Muthukrishnan's \cite{muthukrishnan2002efficient} approach to the document listing problem. We recall it here for completeness.
Let $C$ be an array such that $C[i]$ is the largest $j < i$ where $\TAG'[j] = \TAG'[i]$ (or $-1$ if no such $j$ exists). We want to find all distinct TAG values in the interval $\TAG'[l,r]$.
The first TAG value $\TAG'[i]$ can be determined by a \emph{range-minimum query} (RMQ) on $C[l,r]$, which yields the index $i$ of the minimum in $C[l,r]$; note that $C[i]<l$. Then we recursively consider the sub-intervals $[l,i-1]$ and $[i+1,r]$ of $[l,r]$. Let $[l',r']$ be one of these two intervals. We determine the minimum $C[j]$ in $C[l',r']$ using the RMQ data structure.
If $C[j]\geq l$, then $\TAG[l',r']$ solely contains TAG values that have already been found
and the recursion stops. Otherwise, we output $\TAG'[j]$ and recurse with $[l',j-1]$ and $[j+1,r']$.
The number of range-minimum queries and recursion steps is clearly proportional to the size of the output.

It is undesirable to keep $C$ in memory due to its size.
However, it is possible to equivalently check whether $\TAG'[i]$ has already been output with a global static bit vector where those $\TAG'$ values are marked that have already been output \cite{sadakane2007succinct}. 
Note that this bit vector can be reset in time proportional to the size of the number of set markings \cite{sadakane2007succinct}.

A succinct RMQ data structure that needs $2r_\TAG(1+o(1))$ bits and supports constant-time range-minimum queries can be constructed in linear time \cite{fischer2010optimal}.

\section{Experimental evaluation}
\label{sec:experiments}

We implemented the algorithms and data structures described in this paper in \texttt{C++}.
In particular, our implementation computes the sampled TAG array as described in Sections~\ref{sec:tag_construction}~and~\ref{sec:sampled}
and then computes the sampled TAG index as described in \Cref{sec:sampled} together with the data structure for listing the distinct
TAGs described in \Cref{sec:distinct}.
For the bit vector and $\mathit{rank}$ and $\mathit{select}$ implementations, we used the \emph{Succinct Data Structure Library 2.0} \cite{gog2014sdsl}.
The source code of our implementation is publicly available.\footnote{\url{https://gitlab.com/qwerzuiop/msatag}}

As test data, we use $1000$ human Chromosome~19 haplotypes from \cite{boucher2021phoni} and the corresponding alignment from \cite{olbrich2025alignments}.
All experiments were conducted on a Linux-6.8.0 machine with an Intel Xeon Gold~6338 CPU and 512~GB of RAM.
As the compiler we used GCC~13.3.0.
Currently, our construction algorithm is single-threaded.
We only test our query algorithm with one thread. However, note that concurrent access to our index is trivially possible.

The MSA uses $1.67\cdot 10^8$ gaps and has $5.91\cdot10^{10}$ non-gap characters.
We computed the dollar-$\EBWT$ strands using \texttt{lg} \cite{olbrich2025bwt}.
Ropebwt3 requires that both forward and reverse-complemented strands are in the index, and we of course need to compute the TAG runs with the same index to ensure that the TAG runs match the BWT.
However, we only compute the TAGs for the forward strands. This means that our TAG runs do not cover all positions in the BWT.
Space between two TAG runs is treated like any other TAG run, except marked as ``no TAG available''.

The dollar-$\EBWT$ has $91\,081\,437$ runs, and
there are $137\,981\,814$ runs in our index, $97\,979\,057$ of which have a TAG value.

To test mapping performance, we extracted $10^7$ ``reads'' with 100bp each, chosen uniformly at random from the sequences used for the alignment.
We mutated each base with a probability of $1\%$ and reverse-complemented each read with a probability of $50\%$.
We then used ropebwt3 to find the corresponding MEMs and queried our index with the resulting BWT-interval.
Finally, we projected each TAGs to the first sequence in the data set as a designated linear reference.
ropebwt3 finds $17\,953\,756$ MEMs, which cumulatively correspond to $29\,185\,037$ TAGs (columns in the MSA).

\begin{table}
	\begin{center}
	\begin{tabular}{rrrrr}
	\toprule
	\makecell{Sampling\\rate}
	& \makecell{construction\\time}
	& \makecell{sampled TAG\\runs}
	& \makecell{index\\size}
	& time/TAG \\
	\midrule
	2  &    $213\,\si{\second}$ & $48\,536\,766$ & $455\,\si{\mebi\byte}$ &  $1.74\,\si{\micro\second}$ \\
	4  &    $238\,\si{\second}$ & $23\,800\,106$ & $379\,\si{\mebi\byte}$ &  $2.61\,\si{\micro\second}$\\
	8  &    $300\,\si{\second}$ & $11\,422\,048$ & $340\,\si{\mebi\byte}$ &  $4.11\,\si{\micro\second}$\\
	16 &    $480\,\si{\second}$ &  $5\,244\,736$ & $321\,\si{\mebi\byte}$ &  $6.99\,\si{\micro\second}$\\
	32 &    $992\,\si{\second}$ &  $2\,267\,547$ & $312\,\si{\mebi\byte}$ & $13.29\,\si{\micro\second}$\\
	64 & $2\,002\,\si{\second}$ &  $1\,030\,783$ & $308\,\si{\mebi\byte}$ & $25.31\,\si{\micro\second}$\\
	\bottomrule
	\end{tabular}
	\end{center}
	\caption{For different sampling rates, we list the time for constructing the TAG index (excluding the memory used for $\LF$), the size of the index in memory (excluding $\LF$) and the time per output TAG, averaged over $29\,185\,037$ output TAGs and excluding the time to load the index.
		Constructing the index always needed $2225\,\si{\mebi\byte}$ of memory.
	}
	\label{tab:results}
\end{table}

\Cref{tab:results} shows construction and query performance for varying sampling rates.
The increased construction time with increasing sample rate stems entirely from the construction of the index for reporting the distinct TAG values,
constructing the TAG index always took about $175\,\si{\second}$.
This is because, for constructing this index, we need to access all TAG values, which we do using our TAG index.
Since the average query time of the TAG index is proportional to the sampling rate, a larger sampling rate necessarily slows this down.
By using e.g.\ the non-sampled TAG array or using a more clever algorithm, this could be remedied in the future.
The memory used for constructing the index is roughly the same for all tested sampling rates.

Note that with a sampling rate of $4$, our program takes less time for mapping the MEMs to columns in the MSA with a single thread ($76.0\,\si{\second}$) as ropebwt3 uses for finding the MEMs with $8$ threads ($81.7\,\si{\second}$).

\section{Conclusion}
\label{sec:conclusion}
We described an algorithm that can compute the TAG array of a multiple sequence alignment (MSA) using the $\LF$ function of the BWT
whose working memory is proportional to the number of sequences.
Additionally, we described a TAG index that is able to report the unique tags corresponding to a BWT interval
(i.e., the columns where matches corresponding to the BWT interval occur in the MSA)
in time proportional to the size of the output using standard document listing techniques.
We also gave a non-trivial sampling strategy for the TAG index and showed that our TAG construction algorithm can be adapted
to output just the sampled TAG values.
Finally, we demonstrated experimentally that our construction algorithm and index perform well on real-world data.

Our work enables e.g.\ the efficient mapping of matches in BWT-based indices to a designated reference genome, which programs such as ropebwt3 currently lack \cite{li2024bwt}.
Our techniques could also be used to obtain more effective chaining heuristics in programs using a BWT-based index to find seeds for a seed-and-extend approach (e.g.\ Moni-align \cite{varki2025accurate}).

\bibliography{bib}

\end{document}